\documentclass[preprint,preprintnumbers,showpacs,aps,amssymb]{revtex4}

\usepackage{graphicx}
\usepackage{bm}
\usepackage{amsmath}

\def\calA{{\cal A}}

\def\calO{{\cal O}}
\def\calP{{\cal P}}
\def\calR{{\cal R}}

\def\hbar{{\bar h}}
\def\nbar{{\bar n}}
\def\pbar{{\bar p}}
\def\qbar{{\bar q}}

\def\Ybar{{\bar Y}}

\def\nslash{n\hspace{-2.2mm}/}
\def\nbarslash{\nbar\hspace{-2.2mm}/}
\def\lslash{\ell\hspace{-2.0mm}/}
\def\pslash{p\hspace{-1.8mm}/}

\def\ODY{\langle O_{DY}\rangle}
\begin{document}

\title{Drell-Yan process in soft-collinear effective theory near end-point}
\author{Jong-Phil Lee}
\email{jongphil@korea.ac.kr}
\affiliation{Department of Physics, Korea University, Seoul, 136-701, Korea}

\begin{abstract}
The Drell-Yan process is analyzed in soft-collinear effective theory near the 
end-point region.
It is assumed that the relevant final-state hadron energy $Q(1-z)$ where $z$ is
the momentum fraction transferred to the virtual photon is the typical hadronic
scale $\sim\Lambda$, thus no intermediate scale exists.
It is shown that this setup successfully reproduces the full theory results.
We also discuss the factorized soft Wilson lines for the Drell-Yan process.
\end{abstract}


\maketitle
\section{Introduction}

Hard scattering processes such as Drell-Yan (DY), and deep inelastic scattering
(DIS) are receiving increased attention again nowadays with the development of
effective theories.
\par
There are two important issues for the processes.
One is factorization.
In short, factorization is a separation of long and short distance physics.
Usually transition amplitudes for some processes are given by a factorized 
product or a convolution of hard and soft contributions when the factorization
holds.
The hard part is involved with the short distance physics which can be 
perturbatively calculated.
Long distance physics is encoded with the soft contribution.
In many cases, it is parametrized by the matrix elements of some operators.
Factorization is in general very nontrivial in the full theory \cite{fact}.
But the advent of the soft-collinear effective theory (SCET) \cite{SCET} makes 
it very simple and automatic.
\par
The other is the threshold resummation in the end-point region.
By end-point, we mean $z\to 1$ where $z$ is the usual momentum fraction.
It has been quite well known that there are Sudakov double logarithms
$[\alpha_s\ln^2(1-z)]^n$.
In the end-point region, the large logarithm compensates the small strong
coupling constant \cite{Altarelli}.
The origin of this singularity is the interaction between energetic partons
and soft gluons.
The resummation of these logarithms is necessary and also well studied 
\cite{Catani:1989ne}.
Factorization is again a useful tool in this step \cite{Sterman:1986aj}.
One merit of factorization is that the factorized soft part is universal for
soft emissions.
In \cite{Korchemsky:1992xv,Korchemsky:1993uz}, soft contributions are given 
by the vacuum expectation value of Wilson loops.
The renormalization group evolution of the soft part is governed by the 
cusp anomalous dimensions which originate from the cusp angles in the Wilson
lines.
In position space, the cusp angles are well defined geometrically.
They are responsible for the cusp divergences that give one logarithm of 
$\ln^2$.
The other log comes from the light-cone divergences of cusp angles which are
proportional to $\sim\ln x^2$ where $x$ is a light-like segment 
\cite{Korchemskaya:1992je}.
\par
There have been several works for application of SCET to DIS and DY processes
\cite{Bauer:2002nz,Manohar,CKKL,Ji,Chay:2005rz}.
An interesting kinematic point is $\mu^2\sim Q^2(1-z)\sim Q\Lambda$ where
$\Lambda$ is the hadronic scale.
This so-called ''hard-collinear'' scale \cite{Lunghi:2002ju, Becher} 
appears when soft and collinear particles interact.
The hierarchy $Q^2\gg Q\Lambda\gg\Lambda^2$ ensures the scale separation into
hard, hard-collinear, and soft parts.
By two-step matching \cite{Bauer:2002aj}, one can establish the low energy 
effective theory at $\mu\sim\Lambda$.
Here since the intermediate scale $Q\Lambda$ is still large, it is integrated
out to form the jet functions.
\par
In SCET the factorization is automatic at the operator level.
Especially, the soft gluon effects are compactly factorized in the soft 
Wilson line $Y$.
In many hard scattering processes a universal feature of the soft gluon effects
appears in the proper combination of $Y$s.
Their properties are thoroughly studied in \cite{CKKL}.
\par
There is a slight difference between DIS and DY.
In DIS the final state hadron carries the energy $\sim Q\sqrt{1-z}$ while that
in DY is $\sim Q(1-z)$.
In terms of the moments, the Sudakov double logs are minimized at 
$\mu=Q/\sqrt{N}$ for DIS and $Q=Q/N$ for DY where $N$ is the order of moment.
\par
In DIS, it is quite natural to define a jet function whose momentum scales as
$\sim\sqrt{Q\Lambda}$.
The forward scattering amplitude of DIS is connected with the quark lines and
the intermediate line can be shrunken at the second step of matching by
integrating out the large off-shellness $p_X^2\sim Q^2(1-z)\sim Q\Lambda$, 
which defines unambiguously the jet function \cite{SCET,CKKL}:
\begin{equation}
\langle 0|T\left[W^\dagger\psi(x)~{\bar\psi}W(0)\right]|0\rangle
\equiv
i\int\frac{d^4k}{(2\pi)^4}~e^{-ik\cdot x}J_P(k)\frac{\nslash}{2}~,
\end{equation}
where $W$ is the collinear Wilson line and $J_P$ is the jet function.
On the other hand, in DY there is no final-state hadron at $\calO(\alpha_s^0)$,
nor the hard-collinear scale which defines the intermediate theory
${\rm SCET_I}$ \cite{CKKL}.
\par
In this paper, we do not consider the intermediate scale to separate 
${\rm SCET_I}$ and ${\rm SCET_{II}}$ for DY.
Instead, the full QCD is directly matched onto the final effective theory at
$\mu=Q$.
The end-point region is defined by $1-z\sim\Lambda/Q\ll 1$, so the energy of
final-state hadron $Q(1-z)$ is a small quantity.
This is slightly different from the recent analysis on DY in SCET of \cite{Ji},
where $\mu\sim Q(1-z)$ is a large intermediate scale and two step matching is
implemented.
The main calculation of \cite{Ji}, which is for the soft gluon exchange diagram, 
is very similar to the full QCD analysis of \cite{Altarelli}.
In this work, the same diagram is calculated in a more SCET-based way.
An alternative for the soft gluon effects is the use of soft Wilson lines 
developed in \cite{Bauer:2002ie,CKKL}.
At $\calO(\alpha_s)$ all the soft gluon effects are contained in the factorized
single gluon loop.
It is shown how the new approach gives the same result.
\par 
The paper is organized as follows.
Next Section deals with the basic kinematics and viewpoints on the application
of SCET to DY.
In Section III, the QCD electromagnetic current is matched onto the SCET current 
at $\mu=Q$, and the renormalization of four-quark operator is given.
The factorized soft Wilson line approach appears in Sec.\ IV.
The results of Sec.\ III is reproduced from the time ordered product of soft
Wilson lines. 
In Sec.\ V, the renormalization group evolution and resummation of double 
logarithms are given. 
The cancellation of $\mu$-dependence in the cross section is discussed, and
conclusions are added.

\section{Preliminary}
\subsection{Kinematics}
The center-of-momentum frame of incident hadrons is a natural choice to
describe DY.
For the production of highly virtual photon whose invariant mass is $Q^2>0$, a
parameter $z$ defined by the ratio of $Q^2$ to the invariant mass of partons
governs the kinematics.
Explicitly,
\begin{equation}
(p_1+p_2)^2=2p_1\cdot p_2\equiv s\equiv\frac{Q^2}{z}~,
\end{equation}
where $p_{1,2}$ are the incident partons' momenta. We assume they are massless.
The end-point (or threshold) region is where $z\to 1$.
It is quite convenient to introduce two light-like vectors $n^\mu$ and $\nbar^\mu$
where
\begin{equation}
n^\mu=(1,0,0,1)~,~~~\nbar^\mu=(1,0,0,-1)~.
\end{equation}
They satisfy $n^2=\nbar^2=0$, $n\cdot \nbar=2$.
A momentum $p$ can be decomposed as
\begin{equation}
p^\mu=n\cdot p~\frac{\nbar^\mu}{2}+\nbar\cdot p~\frac{n^\mu}{2}+p_\perp
=(n\cdot p,p_\perp,\nbar\cdot p)=(p^+,p_\perp, p^-)~.
\end{equation}
We choose $p_1(p_2)$ is $\nbar(n)$-collinear:
\begin{eqnarray}
p_1^\mu&=&n\cdot p_1~ \frac{\nbar^\mu}{2}=(n\cdot p_1, 0,0)~,\nonumber\\
p_2^\mu&=&\nbar\cdot p_2~\frac{n^\mu}{2}=(0,0,\nbar\cdot p_2)~.
\end{eqnarray}
Since $s\approx Q^2$ in the end-point region, the photon momentum $q$ can be
set as
\begin{equation}
q^\mu=(Q,0,Q)~,
\end{equation}
where $Q=\sqrt{q^2}$.
Define
\begin{equation}
x_1\equiv\frac{Q^2}{2p_1\cdot q}~,~~~x_2\equiv\frac{Q^2}{2p_2\cdot q}~,
\end{equation}
then we have
\begin{eqnarray}
n\cdot p_1&=&\frac{Q}{x_1}~;~~~p_1=\left(\frac{Q}{x_1},0,0\right)\nonumber\\
\nbar\cdot p_2&=&\frac{Q}{x_2}~;~~~p_2=\left(0,0,\frac{Q}{x_2}\right)~.
\end{eqnarray}
Thus the final state hadron momentum $p_X$ is
\begin{equation}
p_X=p_1+p_2-q=Q\left(\frac{1-x_1}{x_1},0,\frac{1-x_2}{x_2}\right)~,
\end{equation}
and its invariant mass is
\begin{equation}
p_X^2=Q^2\frac{(1-x_1)(1-x_2)}{x_1x_2}
=Q^2\left(\frac{1}{z}+1-\frac{1}{x_1}-\frac{1}{x_2}\right)~,
\end{equation}
where the relation $z=x_1x_2$ is used.
\par
The momenta of mother hadrons are
\begin{eqnarray}
P_1&=&\frac{p_1}{\xi_1}=(\sqrt{S},0,0)~,\nonumber\\
P_2&=&\frac{p_2}{\xi_2}=(0,0,\sqrt{S})~,
\end{eqnarray}
where $\xi_{1,2}$ are momentum fractions and $S$ is the hadronic invariant mass
square.
$s$ and $S$ are related by
\begin{equation}
s=(p_1+p_2)^2=(\xi_1P_1+\xi_2P_2)^2=\xi_1\xi_2S~,
\end{equation}
and
\begin{equation}
z=\frac{Q^2}{s}=\frac{Q^2}{\xi_1\xi_2S}\equiv\frac{\tau}{\xi_1\xi_2}~,
\end{equation}
where $\tau\equiv Q^2/S$.

\subsection{SCET}
SCET is an effective theory for energetic and light particles.
With the kinematics given above, the incident partons are successfully 
described by effective fields of SCET.
We choose $\chi_\nbar$ for $\nbar$-collinear antiquark whose momentum is $p_1$,
and $\xi_n$ for $n$-collinear quark with momentum $p_2$.
Here $\chi_\nbar$ and $\xi_n$ are the standard collinear quark fields of SCET.
\par
The relevant energy scales of DY at the end-point are $Q$ and $Q(1-z)$.
We assume that $1-z\sim\Lambda/Q$ where $\Lambda$ is a typical hadronic scale.
In this case the final state hadron has $p_X^2=Q^2(1-z)^2\sim\Lambda^2$.
It must be compared with the DIS near end-point where the scattered-off parton by
an energetic photon carries $p_{X_{\rm DIS}}^2= Q^2(1-x)\sim Q\Lambda$.
The scale $\sqrt{Q\Lambda}$ is still larger than $\Lambda$, and one introduces
an intermediate theory (${\rm SCET}_{\rm I}$) to integrate out $\sqrt{Q\Lambda}$.
The resulting final theory, ${\rm SCET}_{\rm II}$, is designed to describe only
modes with virtuality of order $\Lambda^2$.
\par
On the other hand, since $Q(1-z)\sim\Lambda$ in DY, there is no intermediate scale
to be integrated out, and thus no intermediate theory is needed.
A direct matching from QCD to SCET at $\mu=Q$ will be enough for DY process at
the end-point.
This approach is quite different from that of \cite{Ji}, where they considered
$Q(1-z)$ as still a large scale compared to $\Lambda$.

\section{Matching and renormalization}

\begin{figure}
\includegraphics{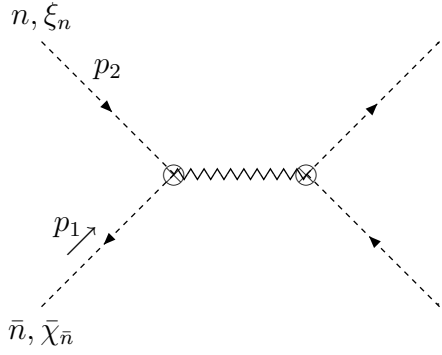}
\caption{\label{tree}Tree diagram of DY process in SCET.}
\end{figure}

At the scale $\mu=Q$, the electromagnetic current in full QCD 
$\qbar_1\gamma^\mu q_2$ is matched onto the SCET current 
${\bar\chi}_\nbar\gamma^\mu\xi_n$.
At tree level, the matching is simple, and the matching coefficient $C(\mu)$ is 
just unity.
Thus the tree level diagram of forward scattering amplitude (Fig.\ \ref{tree}) 
gives
\begin{equation}
i\calA^{\rm tree}=({\bar\xi}_n\gamma^\mu\chi_\nbar)
~\frac{-ig_{\mu\nu}}{(p_1+p_2)^2-Q^2+i0^+}
~({\bar\chi}_\nbar\gamma^\nu\xi_n)~.
\end{equation}
Here the photon is considered as "massive" one whose invariant mass square is
$Q^2$, and $i0^+$ indicates the complex pole position.
Since $(p_1+p_2)^2=2p_1\cdot p_2=s=Q^2/z$, the discontinuity of 
$\calA^{\rm tree}$ is
\begin{equation}
\frac{1}{2\pi i}{\rm Disc.}\calA^{\rm tree}=
({\bar\xi}_n\gamma^\mu\chi_\nbar)({\bar\chi}_\nbar\gamma_\mu\xi_n)
~\frac{1}{Q^2}~\delta(1-z)~.
\end{equation}
The corresponding differential cross section is
\begin{eqnarray}
2s\frac{d\sigma}{dQ^2}&=&\frac{1}{i}{\rm Disc.}\calA^{\rm tree}\nonumber\\
&=&
\frac{2\pi}{Q^2}\delta(1-z)
({\bar\xi}_n\gamma^\mu\chi_\nbar)({\bar\chi}_\nbar\gamma_\mu\xi_n)~.
\end{eqnarray}
This is the same as the full QCD result \cite{Altarelli} up to the overall
normalization.
\par
\begin{figure}
\begin{tabular}{lcr}
\includegraphics{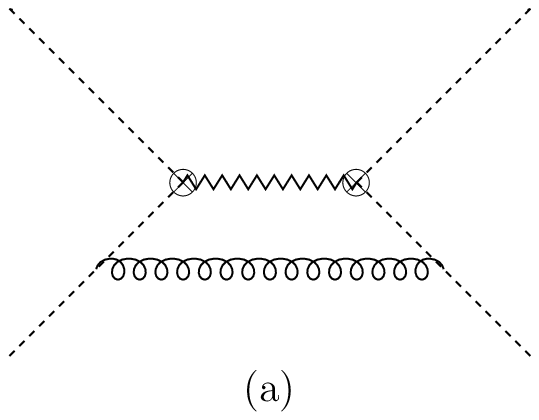}&\includegraphics{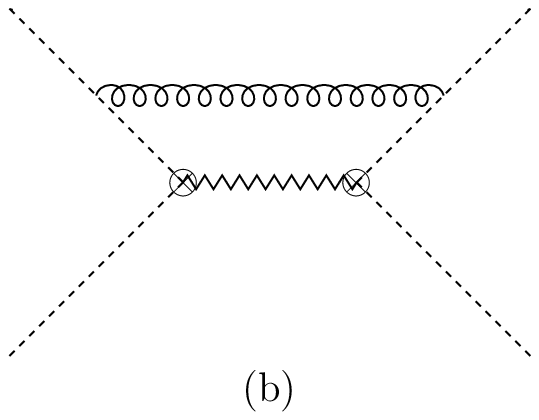}&
\includegraphics{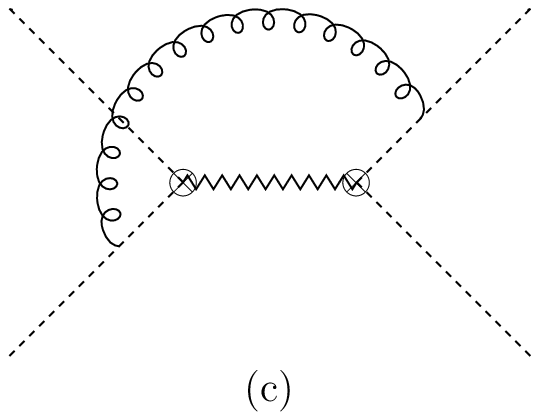}
\end{tabular}
\caption{\label{sg}Soft-gluon one-loop diagrams for DY process in SCET.
Diagram (c) has its mirror image.}
\end{figure}
\begin{figure}
\begin{tabular}{lcr}
\includegraphics{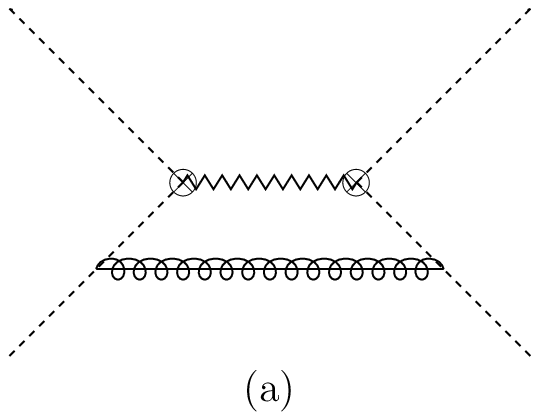}&\includegraphics{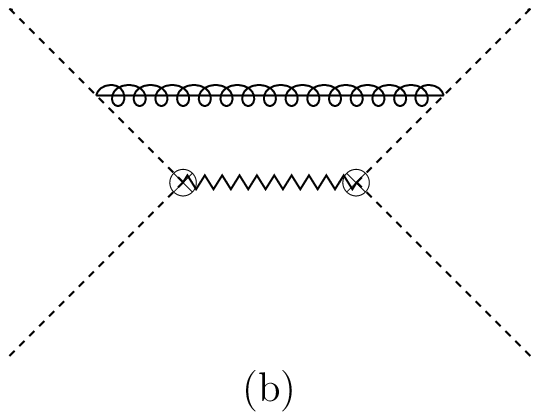}&
\includegraphics{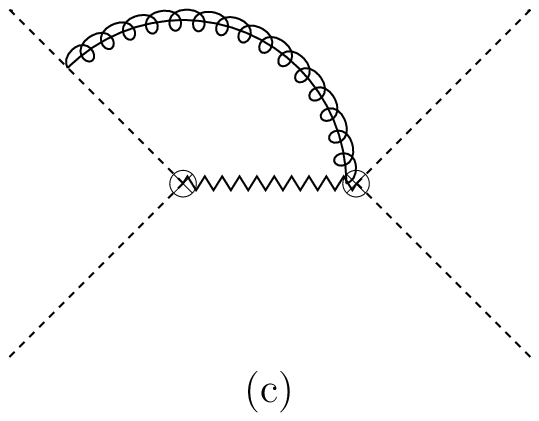}\\
\includegraphics{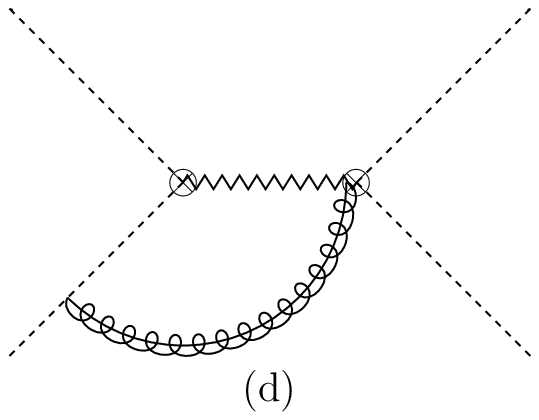}&\includegraphics{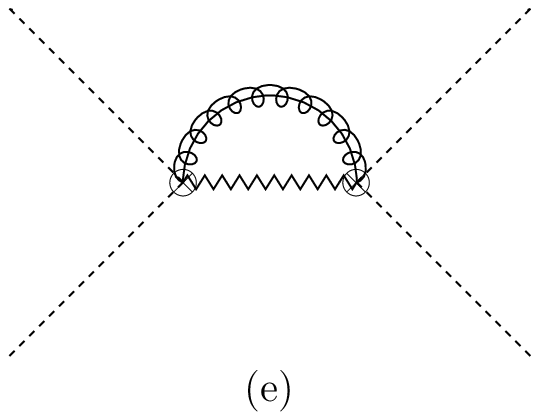}
\end{tabular}
\caption{\label{cg}Collinear-gluon one-loop diagrams for DY process in SCET.
Diagrams (c) and (d) have their mirror images, and (e) contains $n$- and
$\nbar$-collinear gluons.}
\end{figure}

At one-loop level, $C(\mu)$ is also already known in the literature \cite{Manohar}:
\begin{equation}
C(\mu)=1+\frac{\alpha_s}{4\pi}C_F\left(-\ln^2\frac{\mu^2}{Q^2}
 -3\ln\frac{\mu^2}{Q^2}-8+\frac{7\pi^2}{6}\right)~.
\label{Cmu}
\end{equation}
Note that the current matching condition is given by the vertex corrections.
The remaining one-loop diagrams for the forward scattering amplitudes are shown
in Fig.\ \ref{sg} and Fig.\ \ref{cg}.
All these diagrams are responsible for the renormalization of effective 
four-quark operator.
\par
Diagrams Fig.\ \ref{sg} (a) and (b) are proportional to $\sim\nbar^2$ or 
$\sim n^2$, thus vanish.
Nontrivial contribution of the soft gluon comes from Fig.\ \ref{sg} (c).
This diagram is already calculated in \cite{Ji}.
But the calculation is basically not different from the full QCD analysis of 
\cite{Altarelli}.
Here we show how the SCET calculation of Fig.\ \ref{sg} (c) is implemented in
detail and gives the same result as \cite{Altarelli,Ji}.
Explicitly,
\begin{eqnarray}
i\calA_{\rm sg(c)}&=&\int\frac{d^d\ell}{(2\pi)^d}~
{\bar\xi}_n\left(igT^an^\alpha\frac{\nbarslash}{2}\right)\frac{i\nslash}{2}
\frac{\nbar\cdot(p_2+\ell)}{(p_2+\ell)^2+i0^+}\gamma^\mu\chi_\nbar\nonumber\\
&&\times
{\bar\chi}_\nbar\left(igT^b\nbar^\beta\frac{\nslash}{2}\right)
\frac{i\nbarslash}{2}\frac{n\cdot(-p_1-\ell)}{(p_1+\ell)^2+i0^+}\gamma^\nu\xi_n
\nonumber\\
&&\times
\frac{-ig_{\mu\nu}}{(p_1+p_2+\ell)^2-Q^2+i0^+}
\frac{-ig_{\alpha\beta}\delta^{ab}}{\ell^2+i0^+}\nonumber\\
&=&
2g^2C_F({\bar\xi}_n\gamma^\mu\chi_\nbar)({\bar\chi}_\nbar\gamma_\mu\xi_n)
\cdot I_{\rm sg(c)}~,
\end{eqnarray}
where
\begin{eqnarray}
I_{\rm sg(c)}&\equiv&
\int\frac{d^d\ell}{(2\pi)^d}~\frac{1}{\ell^2+i0^+}
~\frac{1}{(p_1+p_2+\ell)^2-Q^2+i0^+}
~\frac{n\cdot(p_1+\ell)}{(p_1+\ell)^2+i0^+}
~\frac{\nbar\cdot(p_2+\ell)}{(p_2+\ell)^2+i0^+}\nonumber\\
&=&
\frac{1}{2}\int\frac{d\ell^+}{2\pi}\frac{d\ell^-}{2\pi}
\frac{d^{d-2}{\vec\ell}_\perp}{(2\pi)^{d-2}}~
~\frac{1}{\ell^+\ell^--{\vec\ell}_\perp^2+i0^+}
~\frac{1}{(\ell^++p_1^+)(\ell^-+p_2^-)-{\vec\ell}_\perp^2-Q^2+i0^+}\nonumber\\
&&\times
~\frac{\ell^++p_1^+}{(\ell^++p_1^+)\ell^--{\vec\ell}_\perp^2+i0^+}
~\frac{\ell^-+p_2^-}{\ell^+(\ell^-+p_2^-)-{\vec\ell}_\perp^2+i0^+}~.
\end{eqnarray}
Since the gluon momentum $\ell$ is soft, 
\begin{equation}
p_1^+,~p_2^-\gg\ell^\pm,~\sqrt{{\vec\ell}_\perp^2}~,
\end{equation}
and the integral becomes
\begin{eqnarray}
I_{\rm sg(c)}&=&
\frac{1}{2}\int\frac{d\ell^+}{2\pi}\frac{d\ell^-}{2\pi}
\frac{d^{d-2}{\vec\ell}_\perp}{(2\pi)^{d-2}}~
~\frac{1}{\ell^+\ell^--{\vec\ell}_\perp^2+i0^+}
~\frac{1}{(\ell^++p_1^+)(\ell^-+p_2^-)-{\vec\ell}_\perp^2-Q^2+i0^+}\nonumber\\
&&\times
~\frac{p_1^+}{p_1^+\ell^-+i0^+} ~\frac{p_2^-}{\ell^+p_2^-+i0^+}~.
\end{eqnarray}
In the photon propagator we keep all the components of $\ell^\mu$ since 
$p_1^+p_2^-\approx Q^2$.
It is convenient to do first the contour integral over $\ell^+$.
All the poles except that in the first term lie in the lower-half plane of
complex $\ell^+$ plane when $\ell^-<0$ and $\ell^-+p_2>0$.
Choosing the contour to cover the upper-half plane,
\begin{equation}
I_{\rm sg(c)}=\frac{i}{2}\int_\theta
\frac{d\ell^-}{2\pi}\frac{d^{d-2}{\vec\ell}_\perp}{(2\pi)^{d-2}}
~\frac{1}{p_2^-{\vec\ell}_\perp^2+p_1^+\ell^-(\ell^-+p_2^-)-Q^2\ell^--i0^+}
~\frac{1}{{\vec\ell}_\perp^2}~,
\end{equation}
where
\begin{equation}
\int_\theta d\ell^-\equiv\int d\ell^-\theta(-p_2^-<\ell^-<0)~.
\end{equation}
Before proceeding, note that the discontinuity of $\calA_{\rm sg(c)}$ is
\begin{equation}
\frac{1}{i}{\rm Disc.}\calA_{\rm sg(c)}
=-2g^2C_F({\bar\xi}_n\gamma^\mu\chi_\nbar)({\bar\chi}_\nbar\gamma_\mu\xi_n)
\frac{1}{i}{\rm Disc.}\left[iI_{\rm sg(c)}\right]~.
\end{equation}
It is quite convenient to take the discontinuity in advance before doing the
integration:
\begin{eqnarray}
\lefteqn{
\frac{1}{i}{\rm Disc.}\left[iI_{\rm sg(c)}\right]}\nonumber\\
&=&
-\frac{1}{2}\int_\theta
\frac{d\ell^-}{2\pi}\frac{d^{d-2}{\vec\ell}_\perp}{(2\pi)^{d-2}}
~\left(\frac{2\pi i}{i}\right)
\delta\left[p_2^-{\vec\ell}_\perp^2+p_1^+\ell^-(\ell^-+p_2^-)-Q^2\ell^-\right]
\frac{1}{{\vec\ell}_\perp}\nonumber\\
&=&
-\frac{1}{2}\int_\theta d\ell^-\left[
\frac{1}{(4\pi)^{d/2-1}}\frac{1}{\Gamma(d/2-1)}~d{\vec\ell}_\perp^2~
\left({\vec\ell}_\perp^2\right)^\frac{d-4}{2}\right]
\frac{1}{p_2^-}
~\delta\left[{\vec\ell_\perp^2-\frac{Q^2-p_1^+(\ell^-+p_2^-)}{p_2^-}\ell^-}\right]
~\frac{1}{{\vec\ell}_\perp^2}\nonumber\\
&=&
-\frac{1}{2p_2^-}\int_\theta d\ell^-
~\frac{1}{(4\pi)^{1-\epsilon}}~\frac{1}{\Gamma(1-\epsilon)}
\left[\frac{Q^2-p_1^+(\ell^-+p_2^-)}{p_2^-}\ell^-\right]^{-1-\epsilon}
\theta\left(\frac{Q^2-p_1^+p_2^-}{p_1^+}<\ell^-<0\right)\nonumber\\
&=&
-\frac{1}{2}\frac{1}{(4\pi)^{d/2-1}}\frac{1}{\Gamma(d/2-1)}
\left(p_1^+p_2^-\right)^{-1-\epsilon}
\left(1-\frac{Q^2}{p_1^+p_2^-}\right)^{-1-2\epsilon}
\int_0^1 dx (1-x)^{-1-\epsilon}x^{-1-\epsilon}~.
\label{disc}
\end{eqnarray}
Since $p_1^+p_2^-=(p_1+p_2)^2=s=Q^2/z$, the whole discontinuity is
\begin{eqnarray}
\label{sgc}
\lefteqn{
\frac{1}{i}{\rm Disc.}\left(\calA_{\rm sg(c)}+{\rm mirror}\right)}\nonumber\\
&=&
\left(\frac{2\pi}{s}\right)O_{DY}
\left(\frac{g^2C_F}{8\pi^2}\right)
\left(\frac{4\pi\mu^2}{Q^2}\right)^\epsilon
~2z^\epsilon(1-z)^{-1-2\epsilon}
~\frac{\Gamma^2(-\epsilon)}{\Gamma(1-\epsilon)\Gamma(-2\epsilon)}~,
\end{eqnarray}
where $O_{DY}=({\bar\xi}_n\gamma^\mu\chi_\nbar)({\bar\chi}_\nbar\gamma_\mu\xi_n)$.
The calculation might have been much easier if the so called cutting rules were applied
from the beginning :
\begin{eqnarray}
\lefteqn{
\frac{1}{i}{\rm Disc.}\left[iI_{\rm sg(c)}\right]}\nonumber\\
&=&
\frac{1}{2}\int\frac{d\ell^+}{2\pi}\frac{d\ell^-}{2\pi}
\frac{d^{d-2}{\vec\ell}_\perp}{(2\pi)^{d-2}}
~(-2\pi i)\delta(\ell^+\ell^--{\vec\ell}_\perp^2)
(-2\pi i)\delta\left[(\ell^++p_1^+)(\ell^-+p_2^-)-{\vec\ell}_\perp^2-Q^2\right]
\nonumber\\
&&\times
\frac{\ell^++p_1^+}{(\ell^++p_1^+)\ell^--{\vec\ell}_\perp}
~\frac{\ell^-+p_2^-}{\ell^+(\ell^-+p_2^-)-{\vec\ell}_\perp}~.
\label{cut}
\end{eqnarray}
It is very easy to see that Eq.\ (\ref{cut}) yields the same result as Eq.\ (\ref{disc}).
\par
Similar manipulations can be done for the collinear gluon exchanges.
Collinear contribution of Fig.\ \ref{cg} (a) is
\begin{eqnarray}
i\calA_{\rm cg(a)}&=&
\int\frac{d^d\ell}{(2\pi)^d}~{\bar\xi}_n\gamma^\mu~\frac{i\nbarslash}{2}~
\frac{n\cdot(p_1+\ell)}{(p_1+\ell)^2+i0^+}~igT^a\nonumber\\
&&\times\left[
\nbar_\alpha+\frac{\gamma^\perp_\alpha(\pslash_1^\perp+\lslash_\perp)}{n\cdot(p_1+\ell)}
+\frac{\pslash_1^\perp\gamma_\alpha^\perp}{n\cdot p_1}
-\frac{\pslash_1^\perp(\pslash_1^\perp+\lslash_\perp)}{n\cdot(p_1+\ell)~n\cdot p_1}
n_\alpha\right]~\frac{\nslash}{2}\chi_\nbar\nonumber\\
&&
\times{\bar\chi}_\nbar igT^b
\left[\nbar_\beta+\frac{\gamma_\beta^\perp\pslash_1^\perp}{n\cdot p_1}
+\frac{(\pslash_1^\perp+\lslash_\perp)\gamma_\beta^\perp}{n\cdot(p_1+\ell)}
-\frac{(\pslash_1^\perp+\lslash_\perp)\pslash_1^\perp}{n\cdot p_1~n\cdot(p_1+\ell)}
n_\beta\right]~\frac{\nslash}{2}\nonumber\\
&&
\times
\frac{i\nbarslash}{2}~\frac{n\cdot(p_1+\ell)}{(p_1+\ell)^2+i0^+}\gamma^\nu
\xi_n~\nonumber\\
&&
\times\frac{-ig_{\mu\nu}}{(p_1+p_2+\ell)^2-Q^2+i0^+}~
\frac{-ig^{\alpha\beta}\delta^{ab}}{\ell^2+i0^+}\nonumber\\
&=&
-g^2C_F({\bar\xi}_n\gamma^\mu\gamma_\perp^\alpha\gamma^\rho\chi_\nbar)
({\bar\chi}_\nbar\gamma_\rho\gamma^\perp_\alpha\gamma_\mu\xi_n)
I_{\rm cg(a)}~,
\end{eqnarray}
where
\begin{equation}
I_{\rm cg(a)}=\int\frac{d^d\ell}{(2\pi)^d}
~\frac{-{\vec\ell}_\perp^2}{d-2}~\frac{1}{\ell^2+i0^+}
~\frac{1}{(p_1+p_2+\ell)^2-Q^2+i0^+}
~\left[\frac{1}{(p_1+\ell)^2+i0^+}\right]^2~.
\end{equation}
The discontinuity of $I_{\rm cg(a)}$ is
\begin{eqnarray}
\lefteqn{\frac{1}{i}{\rm Disc.}\left[iI_{\rm cg(a)}\right]}\nonumber\\
&=&
\frac{1}{2}\int\frac{d\ell^+}{2\pi}\frac{d\ell^-}{2\pi}
\frac{d^{d-2}{\vec\ell}_\perp}{(2\pi)^{d-2}}
~(-2\pi i)\delta(\ell^+\ell^--{\vec\ell}_\perp^2)
(-2\pi i)\delta\left[(\ell^++p_1^+)(\ell^-+p_2^-)-{\vec\ell}_\perp^2-Q^2\right]
\nonumber\\
&&\times
\left(\frac{-{\vec\ell}_\perp^2}{d-2}\right)
\left[\frac{1}{(\ell^++p_1^+)\ell^--{\vec\ell}_\perp^2}\right]^2\nonumber\\
&=&
\left(\frac{2\pi}{s}\right)\frac{1}{8\pi^2}
\left(\frac{4\pi\mu^2}{Q^2}\right)^\epsilon~
\frac{z^\epsilon}{4}(1-z)^{1-2\epsilon}~\frac{\Gamma(\epsilon)}{\Gamma(2-2\epsilon)}~.
\end{eqnarray}
Compared to $I_{\rm sg(c)}$, $I_{\rm cg(a)}$ is suppressed by 
$\sim (1-z)^2\sim\Lambda^2/Q^2$, so we can safely neglect this contribution.
The suppression should occur since Fig.\ \ref{cg} (a) corresponds to the emission of
collinear real gluon in the final state, which is forbidden by the end-point requirements.
That is the reason for the suppression factor $(1-z)^2$.
And Fig.\ \ref{cg} (b) gives the same result as Fig.\ \ref{cg} (a).
It can also be easily found that 
\begin{equation}
\calA_{\rm cg(c)}=-\calA_{\rm cg(d)}~,
\end{equation}
i.e., Figs.\ \ref{cg} (c) and (d) cancel each other.
This is also true for their mirror images.
Diagram Fig.\ \ref{cg} (e) vanishes because it is proportional to 
$\sim n^2(\nbar^2)=0$ for $\nbar(n)$-collinear gluon.
\par
Note that the relevant contribution only comes from Fig.\ \ref{sg} (c).
Using
\begin{equation}
(1-z)^{-1-2\epsilon}=\frac{1}{-2\epsilon}\delta(1-z)
+\left[\frac{(1-z)^{-1-2\epsilon}}{1-z}\right]_+~,
\end{equation}
where "+" denotes the usual plus distribution, we have in $\overline{\rm MS}$ scheme
\begin{eqnarray}
\label{result1}
2s\frac{d\sigma}{dQ^2}&=&
\left(\frac{2\pi}{s}\right) C^2(Q)\langle p\pbar|O_{DY}|p\pbar\rangle
\left(\frac{g^2C_F}{8\pi^2}\right)
\left(\frac{4\pi\mu^2}{Q^2}\right)^\epsilon
~2z^\epsilon(1-z)^{-1-2\epsilon}
~\frac{\Gamma^2(-\epsilon)}{\Gamma(1-\epsilon)\Gamma(-2\epsilon)}\nonumber\\
&=&
\left(\frac{2\pi}{s}\right) C^2(Q)\langle p\pbar|O_{DY}|p\pbar\rangle
\left(\frac{g^2C_F}{8\pi^2}\right)
\left[\frac{2}{\epsilon^2}\delta(1-z)+\frac{2}{\epsilon}\delta(1-z)\ln\frac{\mu^2}{Q^2}
-\frac{4}{\epsilon}\left(\frac{1}{1-z}\right)_+\right.\nonumber\\
&&\left.
+8\left(\frac{\ln(1-z)}{1-z}\right)_+
-4\left(\frac{1}{1-z}\right)_+\ln\frac{\mu^2}{Q^2}
+\delta(1-z)\left(\ln^2\frac{\mu^2}{Q^2}-\frac{\pi^2}{2}\right)\right]~.
\end{eqnarray} 
This is exactly the same as the full QCD result 
\cite{Altarelli,Korchemsky:1993uz}, and \cite{Ji}.

\section{Soft Wilson lines}
In this section we show how the above results can be described in terms of the
soft Wilson line.
\par
The differential cross section of DY process $p_1+p_2\to p_X+q$ is given by
\begin{eqnarray}
2s\frac{d\sigma}{dQ^2}&=&
\sum_X\int\frac{d^3p_X}{(2\pi)^3}\frac{1}{2E_X}
\int\frac{d^3q}{(2\pi)^3}\frac{1}{E_q}
~\langle p\pbar|j^\mu(0)|X\rangle\langle X|j_\mu(0)|p\pbar\rangle
~(2\pi)^4\delta^4(p_1+p_2-p_X-q)\nonumber\\
&=&
\int\frac{d^3q}{(2\pi)^3}\frac{1}{E_q}
\int d^4z e^{-iq\cdot z}\langle p\pbar|j^\mu(z)j_\mu(0)|p\pbar\rangle\nonumber\\
&\equiv&
\int\frac{d^3q}{(2\pi)^3}\frac{1}{E_q}
2{\rm Im}W_{DY}~,
\end{eqnarray}
where
\begin{equation}
W_{DY}\equiv
i\int d^4z e^{-iq\cdot z}\langle p\pbar|T[j^\mu(z)j_\mu(0)]|p\pbar\rangle~,
\end{equation}
and $T$ denotes the time ordering.
The electromagnetic current $j^\mu$ is given by 
\begin{eqnarray}
j^\mu(z)&=&
{\bar\xi}_nWY^\dagger C(\calP_+)\gamma^\mu{\bar Y}{\bar W}^\dagger\chi_\nbar(z)~,
\nonumber\\
j_\mu(0)&=&
{\bar\chi}_\nbar{\bar W}{\bar Y}^\dagger C(\calP_+)\gamma_\mu YW^\dagger\xi_n(0)
\end{eqnarray}
where $\calP_+=\calP+\calP^\dagger$ is the sum of label operators. 
Here the collinear and soft Wilson lines, $W$ and $Y$, are explicitly factorized.
Bars on the Wilson lines mean that the associated collinear direction is $\nbar$.
In what follows, $W$'s are omitted since we are only interested in the soft
Wilson lines.
We adopt the convention of \cite{Arnesen} for $Y$ where the reference 
point of the Wilson line is $-\infty$ uniquely.
This notation is very convenient to check the universality of the cusp angles
in the Wilson lines for various processes as can be seen later.
The soft Wilson line $Y$ is defined by
\begin{equation}
Y=\sum_{\rm perm}\exp\left[-\frac{1}{n\cdot\calR+i0^+}~gn\cdot A_s\right]~,
\end{equation}
where $\calR^\mu=i\partial^\mu$ is the soft momentum operator.
Its Fourier transformation is
\begin{equation}
Y(x)=P\exp\left[ig\int_{-\infty}^x ds~n\cdot A_s(s)\right]~,
\end{equation}
with $P$ being the path ordering.
\par
\begin{figure}
\includegraphics{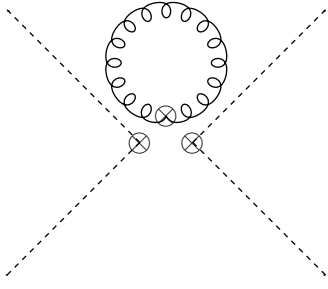}
\caption{\label{sloop}Soft gluon one-loop contribution.}
\end{figure}
\begin{figure}
\includegraphics{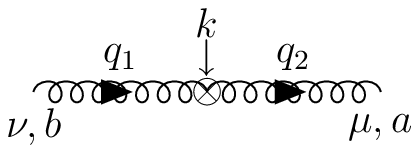}
\caption{\label{sleg}$S_{DY}$ at $\calO(\alpha_s)$.}
\end{figure}

Using $j^\mu(z)=e^{i{\hat p}\cdot z}j^\mu(0)e^{-i{\hat p}\cdot z}$ where 
$\hat p$ is the translation operator, the hadronic matrix element becomes 
($C_+\equiv C(\calP_+)$)
\begin{eqnarray}
\lefteqn{W_{DY}}\nonumber\\
&=&i\int d^4z ~e^{-iq\cdot z}
\langle p\pbar|T\left[e^{i(p_1+p_2)\cdot z}
{\bar\xi}_nY^\dagger C_+\gamma^\mu\Ybar\chi_\nbar(0)
e^{-i{\hat p}\cdot z}\cdot
{\bar\chi}_\nbar\Ybar^\dagger C_+\gamma_\mu Y\xi_n(0)\right]|p\pbar
\rangle\nonumber\\
&=&
\frac{i}{2}\int dz^+ dz^-~(2\pi)^2\delta^2({\vec q}_\perp)\langle p\pbar|
T\left[
\left({\bar\xi}_{n\alpha}\right)_iC_+
\left(Y^\dagger\gamma_{\alpha\beta}^\mu\Ybar\right)_{ij}
\left(\chi_{\nbar\beta}(0)\right)_j\right.\nonumber\\
&&\times\left.
e^{i(p_1^+-q^+-i\partial^+)z^-/2} e^{i(p_2^--q^--i\partial^-)z^+/2}
\left({\bar\chi}_{\nbar\rho}\right)_l C_+
\left(\Ybar^\dagger(\gamma_\mu)_{\rho\lambda}Y\right)_{lm}
\left(\xi_{n\lambda}(0)\right)_m\right]|p\pbar\rangle\nonumber\\
&=&
2i(2\pi)^4\delta^2({\vec q}_\perp)C^2(Q)~
\frac{\delta^{im}}{N_c}
\langle p|T\left[{\bar\xi}_{n\alpha}\xi_{n\lambda}\right]|p\rangle
\gamma^\mu_{\alpha\beta}(\gamma_\mu)_{\rho\lambda}
\frac{\delta^{lj}}{N_c}\langle\pbar|T\left[
{\bar\chi}_{\nbar\rho}\chi_{\nbar\beta}\right]|\pbar\rangle
\nonumber\\
&&\times
\langle 0|T\left[\left(Y^\dagger\Ybar\right)_{ij}
\delta(p_1^+-q^+-i\partial^+)\delta(p_2^--q^--i\partial^-)
\left(\Ybar^\dagger Y\right)_{lm}\right]|0\rangle~,
\end{eqnarray}
where $i,j,l,m(\alpha,\beta,\rho,\lambda)$ are color(Dirac) indices and $N_c$
is color number.
The matrix elements are designed to be color singlet.
We can write
\begin{equation}
W_{DY}
=2(2\pi)^4\delta^2({\vec q}_\perp)C^2(Q)\ODY\cdot iS_{DY}~,
\end{equation}
where
\begin{eqnarray}
\ODY&=&\frac{1}{N_c}
\langle p|T\left[{\bar\xi}_{n\alpha}\xi_{n\lambda}\right]|p\rangle
\gamma^\mu_{\alpha\beta}(\gamma_\mu)_{\rho\lambda}
\langle\pbar|T\left[{\bar\chi}_{\nbar\rho}\chi_{\nbar\beta}\right]|\pbar\rangle~,
\nonumber\\
S_{DY}&=&\frac{1}{N_c}{\rm Tr}\langle 0|T\left[
Y^\dagger\Ybar\delta(p_1^+-q^+-i\partial^+)\delta(p_2^--q^--i\partial^-)
\Ybar^\dagger Y\right]|0\rangle~.
\end{eqnarray}
Here Tr means the trace over color.
The appearance of double delta function in $S_{DY}$ is a peculiar feature of
DY process compared to others \cite{CKKL}.
The nontrivial contribution of $S_{DY}$ comes from one-loop diagram of 
Fig.\ \ref{sloop}.
To make the soft-gluon loop, we first need the Feynman rule for the two soft
gluon legs in Fig.\ \ref{sleg}.
Expanding $Y$ and $\Ybar$, we have
\begin{eqnarray}
\lefteqn{
S_{DY}^{(2)}}\nonumber\\
&=&
\frac{g^2}{N_c}{\rm Tr}\left(T^aT^b\right)\left\{
\left[
\delta(k^++q_2^+-q_1^+)\delta(k^-+q_2^--q_1^-)+\delta(k^+)\delta(k^-)\right]
\frac{\nbar^\mu}{\nbar\cdot q_2-i0^+}~\frac{n^\nu}{\nbar\cdot q_1+i0^+}
\right.\nonumber\\
&&\left.
+\left[\delta(k^+-q_1^+)\delta(k^--q_1^-)
+\delta(k^+-q_2^+)\delta(k^--q_2^-)\right]
\frac{n^\mu}{-n\cdot q_2+i0^+}~\frac{\nbar^\nu}{\nbar\cdot q_1+i0^+}
+(n\leftrightarrow\nbar)\right\}~.\nonumber\\
\end{eqnarray}
The superscript of $S_{DY}^{(2)}$ denotes the power of $g$.
The decoupled soft gluon loop in Fig.\ \ref{sloop} is 
\begin{eqnarray}
S_{DY}^{\rm loop}&=&
\frac{-ig^2}{N_c}\frac{\delta^{aa}}{2}\frac{n\cdot\nbar}{2}
\int\frac{d^d q}{(2\pi)^d}~\frac{1}{q^2+i0^+}\left[
2\delta(p_X^+)\delta(p_X^-)\frac{1}{q^+-i0^+}\frac{1}{q^-+i0^+}\right.\nonumber\\
&&
\left.
-\delta(p_X^+-q^+)\delta(p_X^--q^-)\left(\frac{1}{q^++i0^+}\frac{1}{q^--i0^+}
+\frac{1}{q^+-i0^+}\frac{1}{q^-+i0^+}\right)+(q^\pm\to-q^\pm)\right]~.\nonumber\\
\end{eqnarray}
At this stage, recall that $d\sigma\sim2{\rm Im}W_{DY}$, and
\begin{equation}
{\rm Im}W_{DY}=2(2\pi)^4\delta^2({\vec q}_\perp)C^2(Q)\ODY
{\rm Im}\left(iS_{DY}\right)~.
\end{equation}
Terms proportional to $\delta(p_X^+)\delta(p_X^-)$ are zero in pure
dimensional regularization.
The remaining contributions give
\begin{eqnarray}
{\rm Im}\left(iS_{DY}^{\rm loop}\right)&=&
-2\times g^2C_F~\frac{1}{2}~{\rm Im}\left[
\int\frac{d\ell^+}{2\pi}\frac{d\ell^-}{2\pi}
\frac{d^{d-2}{\vec\ell}_\perp}{(2\pi)^{d-2}}~
\frac{\delta(p_X^+-\ell^+)\delta(p_X^--\ell^-)}{\ell^+\ell^--{\vec\ell}_\perp^2+i0^+}
~\frac{2}{\ell^+\ell^-}\right]\nonumber\\
&=&
-2\times g^2C_F~\frac{1}{2(2\pi)^2}
\int\frac{d^{d-2}{\vec\ell}_\perp}{(2\pi)^{d-2}}~
(-\pi)\delta(p_X^+p_X^--{\vec\ell}_\perp^2)~\frac{2}{p_X^+p_X^-}\nonumber\\
&=&
\frac{g^2C_F}{8\pi^2}~\frac{(4\pi)^\epsilon}{\Gamma(1-\epsilon)}~
\left(p_X^+p_X^-\right)^{-1-\epsilon}~.
\end{eqnarray}
Here the sign-flipped terms yield the same result, and that is the reason of
the overall factor 2.
The differential cross section is
\begin{eqnarray}
2s\frac{d\sigma}{dQ^2}&=&
\int\frac{d^3{\vec q}}{(2\pi)^3}\frac{1}{2E_q}~2{\rm Im}W_{DY}\nonumber\\
&=&
\int\frac{d^3{\vec q}}{(2\pi)^3}\frac{1}{2E_q}~
4(2\pi)^4\delta^2({\vec q}_\perp)C^2(Q)\ODY~
\frac{g^2C_F}{8\pi^2}~\frac{(4\pi\mu^2)^\epsilon}{\Gamma(1-\epsilon)}~
\left(p_X^+p_X^-\right)^{-1-\epsilon}~.
\end{eqnarray}
The relevant phase space integral is
\begin{eqnarray}
\lefteqn{
\int\frac{d^3{\vec q}}{(2\pi)^3}\frac{1}{2E_q}~(2\pi)^4\delta^2({\vec q}_\perp)
\left(p_X^+p_X^-\right)^{-1-\epsilon}}\nonumber\\
&=&
\int\frac{d^4 q}{(2\pi)^3}\delta(q^2-Q^2)~(2\pi)^4\delta^2({\vec q}_\perp)
\left(p_X^+p_X^-\right)^{-1-\epsilon}\nonumber\\
&=&
(2\pi)\int d^4p_X~\delta\left[(p_1+p_2-p_X)^2-Q^2\right]~\delta^2({\vec p}_X^\perp)
\left(p_X^+p_X^-\right)^{-1-\epsilon}\nonumber\\
&=&
(2\pi)\frac{Q^\epsilon}{2}(p_2^-)^{-1-\epsilon}
\int_0^{\frac{s-Q^2}{p_2^-}}~dp_X^+(p_X^+)^{-1-\epsilon}
\left(\frac{s-Q^2}{p_2^-}-p_X^+\right)^{-1-\epsilon}~,
\end{eqnarray}
where we used the fact that $p_1^+-p_X^+=Q$ and $p_X^-=Q(1-x_2)/x_2>0$.
Integral over $p_X^+$ gives the beta function, and the final result is 
\begin{equation}
2s\frac{d\sigma}{dQ^2}=
\left(\frac{2\pi}{s}C^2(Q)\ODY\right)
\left(\frac{g^2C_F}{8\pi^2}\right)
\left(\frac{4\pi\mu^2}{Q^2}\right)^\epsilon
2z^\epsilon(1-z)^{-1-2\epsilon}
\frac{\Gamma^2(-\epsilon)}{\Gamma(1-\epsilon)\Gamma(-2\epsilon)}~.
\end{equation}
This is exactly Eq.\ (\ref{result1}), since 
$\langle p\pbar|O_{DY}|p\pbar\rangle=\ODY$.
\par
\begin{figure}
\includegraphics{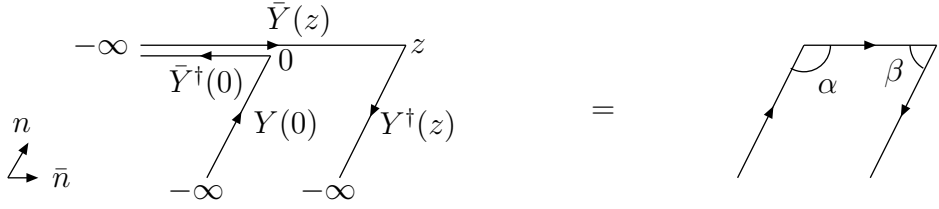}
\caption{\label{DYWilson}Soft Wilson lines $S_{DY}$.}
\end{figure}

The procedure implemented in this section can be explained geometrically
by the universal cusp angle of the Wilson lines.
Figure \ref{DYWilson} shows $S_{DY}$.
The resulting Wilson line (right one in Fig.\ \ref{DYWilson}) is just that of 
DIS \cite{Korchemsky:1992xv,CKKL}.
Thus the cusp angles are common in DY and DIS.
Before cutting, the Wilson lines have two distinctive cusp angles,
$\alpha$ and $\beta$.
They are responsible for the cusp anomalous dimensions.
What is different is the involved energy, $Q^2(1-z)^2\sim\Lambda^2$ for DY 
and $Q^2(1-x)\sim Q\Lambda$ for DIS.
If we express $S_{DY}$ in position coordinates, $\alpha$ and $\beta$ have
different $i\epsilon$ prescriptions for the corresponding eikonal lines.
Taking a cut means dagger operation for $\beta$, resulting in two identical
cusp angles \cite{Korchemsky:1992xv}.
\par
The above picture is slightly different from \cite{CKKL}.
Within the convention of \cite{CKKL} for the soft Wilson lines, there are two 
reference points, $\pm\infty$.
In this convention, the hermiticity of the collinear spinor
$\xi_n$ is violated and a nontrivial factor 
$Y_\infty\equiv P\exp\left[ig\int_{-\infty}^\infty ds~n\cdot A_s(s)\right]$
is involved \cite{Arnesen}.
The resulting Wilson lines for DIS and DY are quite different from those in 
this work or other literature, exhibiting two {\em identical} cusp angles 
(see Figs.\ 3(c) and 4(b) of \cite{CKKL}).
Any kinds of cut operations are not needed in the soft Wilson lines afterward.
The reason is that in the framework of \cite{CKKL}, the soft Wilson lines do
not have discontinuity at least at one loop level; 
the imaginary part of the hadronic tensor comes solely from the jet function 
from the construction.
Also, the intermediate soft Wilson lines are chosen such that the relative 
extra factor $Y_\infty$ between the two conventions does not appear, which
results in the same physical observables.
A more general discussion about the choice of convention can be found in
\cite{Arnesen}.
\section{Discussions and conclusions}

The finite part of soft gluon contributions to $d\sigma$ is (including tree level)
\begin{eqnarray}
F_{DY}&\equiv& 
2s\left(\frac{d\sigma}{dQ^2}\right)_{\rm finite}/
\left(\frac{2\pi}{s}C^2(Q)\ODY\right)
\nonumber\\
&=&
\delta(1-z)+\left(\frac{\alpha_sC_F}{2\pi}\right)\left[
8\left(\frac{\ln(1-z)}{1-z}\right)_+
-4\left(\frac{1}{1-z}\right)_+\ln\frac{\mu^2}{Q^2}\right.\nonumber\\
&&\left.
+\delta(1-z)\left(\ln^2\frac{\mu^2}{Q^2}-\frac{\pi^2}{2}\right)\right]~.
\end{eqnarray}
The moment of $F_{DY}$ is
\begin{eqnarray}
F_N&=&\int_0^1 dz~z^{N-1}F_{DY}\nonumber\\
&=&
1+\left(\frac{\alpha_sC_F}{2\pi}\right)\left[
8\sum_{i=1}^{N-1}\frac{1}{i}\sum_{j=1}^i\frac{1}{j}
+4\sum_{i=1}^{N-1}\frac{1}{i}\ln\frac{\mu^2}{Q^2}
+\ln^2\frac{\mu^2}{Q^2}-\frac{\pi^2}{2}\right]\nonumber\\
&\rightarrow&
1+\left(\frac{\alpha_sC_F}{2\pi}\right)\left[
4\ln^2\left(\frac{\mu}{Q}\frac{N}{N_0}\right)+\frac{\pi^2}{6}\right]~,
\label{FN}
\end{eqnarray}
for large $N$.
Here $N_0=e^{-\gamma_E}$ where $\gamma_E$ is the Euler number.
In our picture, the large scale $\mu\sim Q$ determines lower and upper bounds for
hard and soft physics, respectively \cite{Korchemsky:1993uz}.
\par
The renormalization group equation (RGE) for $F_N$ is
\begin{equation}
\left(\mu\frac{\partial}{\partial\mu}
+\beta(g)\frac{\partial}{\partial g}\right)\ln F_N
=\left(\frac{\alpha_sC_F}{\pi}\right)2\ln\left(\frac{\mu^2}{Q^{\prime 2}}\right)~,
\end{equation}
where $Q'=QN_0/N$.
The solution of RGE is simply
\begin{equation}
F_N(\mu)=F_N(Q')\exp\left[\int_{Q'^2}^{\mu^2}\frac{d\nu^2}{\nu^2}~
\left(\frac{\alpha_s(\nu^2)C_F}{\pi}\right)
\ln\left(\frac{\nu^2}{Q'^2}\right)\right]~.
\end{equation}
Note that $\mu\ge Q'\sim\Lambda$.
This is quite different from \cite{Ji} where $Q'$ is still large.
\par
The counter term of the SCET current is \cite{Manohar,Ji}
\begin{equation}
{\rm c.t.}=\left(\frac{2\pi}{s}\ODY\right)\left(\frac{g^2C_F}{8\pi^2}\right)
\left(-\frac{1}{\epsilon^2}-\frac{3}{2\epsilon}
-\frac{1}{\epsilon}\ln\frac{\mu^2}{Q^2}\right)\delta(1-z)~.
\label{ct}
\end{equation}
Including Eq.\ (\ref{ct}), the divergent part of $d\sigma$ is
\begin{equation}
2s\left(\frac{d\sigma}{dQ^2}\right)_{\rm div}+2\times{\rm c.t.}=
\left(\frac{2\pi}{s}\ODY\right)\left(\frac{\alpha_sC_F}{-2\pi\epsilon_{IR}}\right)
2P_{q\to q}(z)~,
\end{equation}
where
\begin{equation}
P_{q\to q}(z)\equiv 
C_F\left[\frac{3}{2}\delta(1-z)+2\left(\frac{1}{1-z}\right)_+\right]~,
\end{equation}
is the Altarelli-Parisi (AP) kernel.
The infrared divergence is cancelled by the renormalized effective quark operator
$O_n={\bar\xi}_n\Gamma\xi_n$ where $\Gamma$ is some gamma matrix.
The corresponding renormalization constant is given by \cite{Manohar}
\begin{equation}
Z_n(x)=\delta(1-x)+\frac{\alpha_s}{2\pi\epsilon}~P_{q\to q}(x)~.
\end{equation}
The complete expression of $d\sigma$, including the SCET current counter terms, is
\begin{eqnarray}
\lefteqn{
\left[2s\left(\frac{d\sigma}{dQ^2}\right)+2\times{\rm c.t.}\right]}
\nonumber\\
&=&
\left(\frac{2\pi}{s}\right)C^2(\mu)
\left[\delta(1-z)+\frac{\alpha_s}{2\pi}
\left(\frac{1}{\epsilon}+\ln\frac{\mu^2}{Q^2}\right)2P_{q\to q}(z)\right]\ODY
\cdot G(z)\nonumber\\
&\equiv&
2s\left(\frac{d\sigma}{dQ^2}\right)_{\rm NLO}~,
\end{eqnarray}
where $G(z)-1$ is the soft contribution,
\begin{equation}
G(z)=1+\left(\frac{4\pi\mu^2}{Q^2}\right)^\epsilon
\left(\frac{\alpha_sC_F}{2\pi}\right)
~2z^\epsilon(1-z)^{-1-2\epsilon}~
\frac{\Gamma^2(-\epsilon)}{\Gamma(1-\epsilon)\Gamma(-2\epsilon)}~.
\label{full}
\end{equation}
Here we specified $C=C(\mu)$ to show how the $\mu$-dependence of Eq.\ (\ref{full})
is cancelled.
Plugging Eqs.\ (\ref{Cmu}) and (\ref{result1}) into (\ref{full}), we have
\begin{eqnarray}
\lefteqn{2s\left(\frac{d\sigma}{dQ^2}\right)_{\rm NLO}/
\left(\frac{2\pi}{s}\right)\ODY}\nonumber\\
&=&
\delta(1-z)+\frac{\alpha_sC_F}{2\pi}\left\{
8\left(\frac{\ln(1-z)}{1-z}\right)_++\delta(1-z)\left(-8-\frac{\pi^2}{3}\right)
\right.\nonumber\\
&&\left.
-\left[4\left(\frac{1}{1-z}\right)_++3\delta(1-z)\right]\ln\frac{\mu^2}{Q^2}
+\frac{2P_{q\to q}(z)}{C_F}\ln\frac{\mu^2}{Q^2}\right\}\nonumber\\
&=&
\delta(1-z)+\frac{\alpha_sC_F}{2\pi}\left\{
8\left(\frac{\ln(1-z)}{1-z}\right)_++\delta(1-z)\left(-8-\frac{\pi^2}{3}\right)
\right\}~.
\end{eqnarray}
Note that the $\mu$-dependence of $C(\mu)$ combined with that of $G(z)$ is cancelled
by the AP kernel of the quark operator \cite{Korchemsky:1992xv}.
\par
In \cite{Sterman:1986aj}, factorization and resummation of Sudakov logs are analyzed
in a general form.
They also use the terminology ''jet function'' but the meaning is slightly different
from that in SCET.
In \cite{Sterman:1986aj}, the ''jet'' functions of DY scale as $Q/\mu N$ where 
$N$ is the order of moments.
These are the usual parton distribution functions, and correspond to $\ODY$ in
this work.
The scaling behavior of the ''jet'' functions is the same as that of the ''soft 
function'', while the hard function scales as $Q^2/\mu^2$.
On the other hand, the ''jet'' function of DIS scales as $Q/\mu N^{1/2}$ which
exactly corresponds to the SCET jet function.
The present work does not introduce the SCET jet functions in DY since no 
intermediate energy scale is assumed.
The imaginary part of $W_{DY}$ appears only in the soft Wilson lines.
\par
Note that there is no scaling like $\sim N^{-1/2}$ in DY in full QCD.
This is because no intermediate scale $\sim\sqrt{Q\Lambda}$ is assumed.
The characteristic scaling of DY is $N^{-1}$ which leads to a familiar
correspondence $\Lambda/Q\sim 1-z\sim 1/N$ in QCD.
In DIS where a large intermediate scale can exist, the Bjorken variable $x$ 
scales at the end-point as
$1-x\sim\Lambda/Q\sim1/N\gg\Lambda^2/Q^2$, where the final state hadron has
a large off-shellness, $p_X^2=Q^2(1-x)\sim Q\Lambda\gg\Lambda^2$
\cite{Korchemsky:1993uz,Manohar}.
In \cite{Ji}, a similar relation holds, $1-z\sim 1/N$, but in this case
$1-z\gg\Lambda/Q$ to make the intermediate scale $Q(1-z)$ large enough.
Thus the moment scales as $\Lambda/Q\ll 1/N$ in \cite{Ji}, which might not be
true for sufficiently large $N$.
In this work, $1-z\sim\Lambda/Q\sim 1/N$ just like in QCD, as can be seen in 
Eq.\ (\ref{FN}).
Consequently, $1-z$ behaves much better than that of \cite{Ji} for large $N$,
and our scaling behavior is consistent with the literatures.
\par
In summary, DY process is analyzed in SCET near end-point.
With the scaling of $Q(1-z)\sim\Lambda$, full QCD is directly matched onto 
SCET at $\mu=Q$.
Previously well known results are successfully reproduced.
In addition, it is shown that the soft Wilson line approach ensures that the 
factorization of soft gluon effects is automatic, and a single gluon loop
diagram gives the same soft gluon corrections at $\calO(\alpha_s)$.
The structure of RGE is rather simple because there is no intermediate scale.
It is also discussed how the scale dependence vanishes in the cross section.


\end{document}